\documentclass[aps,prl,epsfigure,twocolumn,showpacs]{revtex4}
\usepackage{dcolumn}
\usepackage{bm}

\usepackage{graphicx}
\usepackage{amsmath}
\usepackage{latexsym}
\usepackage{amsfonts}
\usepackage{amssymb}
\usepackage{array}
\usepackage{epsfig}

\newcommand{\one}{\mbox{$1 \hspace{-1.0mm}  {\bf l}$}}

\newcommand{\nbar}{\overline{n}}

\begin{document}


\title{Creating and probing macroscoping entanglement with light}
\author{M. Paternostro$^{1}$, D. Vitali$^2$, S. Gigan$^{3,4}$, M. S. Kim$^1$, C. Brukner$^{1,4}$, J. Eisert$^5$, M. Aspelmeyer$^{1,4}$}

\affiliation{$^1$School of Mathematics and Physics, Queen's University, Belfast BT7 1NN, United Kingdom\\
$^2$Dipartimento di Fisica, Universita di Camerino, I-62032 Camerino (MC), Italy\\
$^3$Institute for Quantum Optics and Quantum Information (IQOQI), Boltzmanngasse 3, A-1090 Vienna, Austria\\
$^4$Institute for Experimental Physics, University of Vienna, Boltzmanngasse 5, A-1909 Vienna, Austria\\
$^5$Blackett Laboratory, Imperial College London, Prince Consort Road, London SW7 2BW, UK\\
\& Institute for Mathematical Sciences, Imperial College London, Prince's Gardens, London SW7 2PE, UK}

\date{\today}

\begin{abstract}

We describe a scheme showing signatures of macroscopic optomechanical entanglement generated by radiation pressure in a cavity system with a massive movable mirror. The system we consider reveals genuine multipartite entanglement. We highlight the way the entanglement involving the inaccessible massive object is unravelled, in our scheme, by means of field-field quantum correlations.

\end{abstract}


\maketitle


Entanglement is currently at the heart of physical investigation not just because of its critical role in setting the mark between the
classical and quantum world but also because of its exploitability in many quantum information tasks~\cite{libri}. So far, theoretical and experimental endeavors have been directed towards the demonstration of entanglement between microscopic systems, mainly for the purposes of information processing and manipulation~\cite{libri}. Nevertheless, the possibility of observing non-classical correlations in systems of macroscopic objects and in situations close to be {\it classical} is very appealing and efforts have been made along this direction~\cite{varie}.
The interest has been also extended to micro- and nanomechanical oscillators, which have been shown to be highly controllable and represent natural candidates for quantum limited measurements, quantum state engineering and for testing decoherence theories~\cite{varie2,david}. This inspired us to study an optomechanical device as a macroscopic system with readily achievable non-classicality. 
%

In spite of these exciting progresses, it is in general still difficult to infer the quantum properties of a macroscopic object. For the case at hand, the properties of the mirror are not directly accessible and one has to design strategies to infer its dynamics~\cite{david,io}. Here we discuss a protocol to unravel the quantum correlations established in a cavity with a moving mirror. Our scheme uses an ancillary cavity field
interacting with the optomechanical device. We treat the whole field-mirror-field system as intrinsically {\it tripartite} and investigate the behavior of entanglement between the subparties. We see signatures of mirror-field entanglement in the quantum correlations between the two cavity fields, which can be quantified through a simple reconstruction algorithm. Our study, together with weak assumptions concerning the underlying model, paves the way to probe information about a system 
by getting entangled with it. 
%
%
We provide an assessment about entanglement inference in a situation of current experimental interest~\cite{nature} where one of the parties is not accessible.

\begin{figure}[t]
\psfig{figure=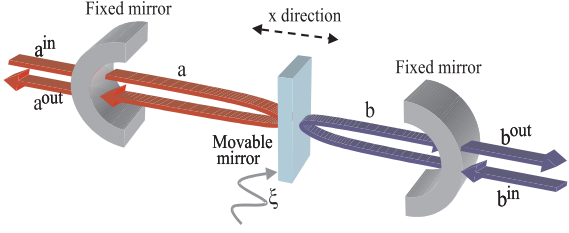,width=7.5cm,height=3.0cm}
\caption{Sketch of the system considered. Fields $a$ and $b$ interact with a movable mirror. The two cavities are driven by input fields with power $P_{a,b}$. Input (output) fields are indicated as $\hat{j}^{in}$ ($\hat{j}^{out}$) with $j=a,b$. $\hat{\xi}$ describes the Brownian motion of the mirror at temperature $T$.}
\label{readout}
\end{figure}

{\it The model. --}  The setup we consider consists of two optical cavities labelled $a$ and $b$, each in a Fabry-Perot configuration, sharing a movable mirror. The input mirrors of the cavities are assumed to be fixed and each cavity is driven by an external field of frequency $\omega_{lj}$, input power $P_j$ ($j=a,b$) and coupling strength $E_j$. The system is sketched in Fig.~\ref{readout}. The field of cavity $j$, locked at the frequency $\omega_{j}\simeq{\omega}_{lj}$, is described by the annihilation (creation) operator $\hat{j}$ ($\hat{j}^\dag$). In terms of field quadratures, $\hat{j}=(\hat{x}_j+i\hat{y}_j)/\sqrt{2}$. The mirror is modelled as a single bosonic mode with frequency $\omega_m$ and mass $\mu$. It undergoes quantum Brownian motion due to its contact to a bath at temperature $T$ given by background modes. For standard Ohmic noise characterized by a coupling stength $\gamma_m$, this leads to a non-Markovian correlation function of the associated noise operator $\hat\xi$ of the form $\langle\hat\xi(t)\hat\xi(t')\rangle=(\gamma_m/\omega_m)\int{\omega}e^{-i\omega(t-t')}[1+\coth(\beta\omega/2)]d\omega/2\pi$ (with $\beta=\hbar/k_BT$ and $k_B$ the Boltzmann constant) \cite{giovannettivitali}. As shown, under realistic conditions of weak coupling of the mirror to the environment, a Markovian description can be gained. In this setting the mirror motion is damped at a rate $\gamma_m$.

In a frame rotating at the frequency of the lasers, the energy of the system is written as 
\begin{equation}
\label{hamiltoniana}
\hat{H}=\frac{\hbar\omega_m}{2}(\hat{p}^2\!+\!\hat{q}^2)+\hbar\sum_{j=a,b}\left[(\Delta_{0j}-\tilde{G}_{0j}\hat{q})\hat{j}^\dag{\hat{j}}+i{E}_j(\hat{j}^\dag-\hat{j})\right].
\end{equation}
Here, $\hat{p},\,\hat{q}$ are the mirror dimensionless quadrature operators, $\tilde{G}_{0j}=(-1)^{\delta_{jb}}{G}_{0j}$ with 
$G_{0j}=({\omega_{j}}/{{\ell}_j})
({{\hbar}/({\mu\omega_m})})^{1/2}$ 
the optomechanical coupling rate between the mirror and the $j$-th cavity (length ${\ell}_j$), $\Delta_{0j}=\omega_j-\omega_{lj}$ and $|E_{j}|=({{2\kappa_jP_j}/({\hbar\omega_{lj}})})^{1/2}$ 
with $\kappa_j$ the $j$-th cavity decay rate. 

In order to study the evolution of the system we refer to the Heisenberg picture. The intrinsically open dynamics at hand
is well described by a set of Langevin equations 
obtained considering the {fluctuations} around the mean values of the operators in the problem 
and neglecting any resulting non-linear term. This is a well-established tool allowing for the exact reconstruction of the quantum statistical properties of the system, as far as the fluctuations of the operators are small compared to the mean values~\cite{fabre}.
By defining the equilibrium position of the mirror $q_s=\sum_j{\tilde G}_{0j}|\alpha_{s,j}|^2/\omega_m$, the stationary amplitudes of the intracavity fields 
$\alpha_{s,j}=|E_{j}|/({\kappa^2_j+\Delta^2_j})^{1/2}$~\cite{commentophase} and the effective detunings $\Delta_{j}=\Delta_{0j}-{\tilde G}_{0j}q_s$, the linearized Langevin equations for fluctuations read 
\begin{eqnarray}
\label{langevin3B}
\partial_t\delta\hat{q}&=&\omega_m\delta\hat{p},\\
\partial_t\delta\hat{p}&=&
-\omega_m\delta\hat{q}-\gamma_m\delta\hat{p}+\sum_{j=a,b}\sqrt{2}\tilde{G}_{0j}\alpha_{s,j}\delta\hat{x}_j+\hat\xi,
\nonumber\\
\partial_t\delta\hat{x}_j&=&-\kappa_j\delta\hat{x}_j+\Delta_{j}\delta\hat{y}_j+(-1)^{\delta_{jb}}\delta\hat{X}^{in}_{j},\nonumber\\
\partial_t\delta\hat{y}_j&=&-\kappa_j\delta\hat{y}_j-\Delta_{j}\delta\hat{x}_j+\sqrt{2}\tilde{G}_{0j}\alpha_{s,j}\delta\hat{q}+(-1)^{\delta_{jb}}\delta\hat{Y}^{in}_{j}.\nonumber
\end{eqnarray}

{\noindent}{W}e have introduced the quadrature operators associated with the input noise to the cavities $\delta{\hat Q}^{in}_j=\sqrt{2\kappa_j}\delta{\hat q}^{in}_{j},\,(Q=X,Y;\,q=x,y)$ and the operator $\hat{\xi}$ accounting for the zero-mean Brownian noise. The input noise is correlated as $\langle\delta\hat{a}_{in,j}(t)\delta\hat{a}_{in,k}^{\dag}(t')\rangle=\delta_{jk}\delta(t-t')$ with $\delta\hat{a}_{in,j}=(\delta\hat{x}_{in,j}+i\delta\hat{y}_{in,j})/\sqrt{2}$. 
Moreover, for $\gamma_m\ll\omega_m$ we have $\langle\hat\xi(t)\hat\xi(t')+\hat\xi(t')\hat\xi(t)\rangle\propto{\delta(t-t')}$~\cite{ben}.    
The linearity of Eqs.~(\ref{langevin3B}) preserves the Gaussian
character of the system. We can define the vector ${\hat{\bf f}}^T=(\delta\hat{x}_{a},\delta\hat{y}_{a},\delta\hat{x}_{b},\delta\hat{y}_{b},\delta\hat{q},\delta\hat{p})$, the kernel ${\mathbf K}$ ($G_{j}\!=\!\sqrt{2}\alpha_{s,j}G_{0j}$)
\begin{equation}
\label{kernel3B}
{\mathbf K}=\left[
\begin{matrix}
-\kappa_a&\Delta_a&0&0&0&0\\
-\Delta_a&-\kappa_a&0&0&G_a&0\\
0&0&-\kappa_b&\Delta_b&0&0\\
0&0&-\Delta_b&-\kappa_b&-G_b&0\\
0&0&0&0&0&\omega_m\\
G_a&0&-G_b&0&-\omega_m&-\gamma_m\\
\end{matrix}\right]
\end{equation}
and the noise correlation matrix $\langle\hat{n}_p(t)\hat{n}_q(t')\!+\!\hat{n}_q(t')\hat{n}_p(t)\rangle/2\!=\!{\mathbf N}_{pq}\delta(t-t')$ associated with the noise vector $\hat{\bf n}^T(t)=(\delta\hat{X}^{in}_{a},\delta\hat{Y}^{in}_{a},-\delta\hat{X}^{in}_{b},-\delta\hat{Y}^{in}_{b},0,\hat\xi)$. Here
${\mathbf N}=\!{\kappa_a}\one_2\oplus{\kappa_b}\one_{2}\!\oplus\!{\bm \Xi}~{\rm where}~{\bm \Xi}\simeq{\rm Diag}[0,\gamma_m(2\nbar+1)]
$
with $\nbar=(e^{\beta{\omega_m}}-1)^{-1}$ and $\one_2$ the $2\times{2}$ identity matrix.
Eqs.~(\ref{langevin3B}) are solved as
$\hat{\bf f}(t)=e^{{\mathbf K}t}\hat{\bf f}(0)+\int^{t}_{0}d{\tau}e^{{\mathbf K}\tau}\hat{\bf n}(t-\tau)$.

We aim at studying the entanglement properties of the steady state, which is guaranteed to exist if
the real parts of the eigenvalues of ${\mathbf K}$ are negative.
For the purposes of our work it is sufficient to state that this requirement is equivalent to the positivity 
of two functions, named $C_1$ and $C_2$, the latter of
which can be constructed as described in Ref.\ ~\cite{RH}.
\begin{figure}[b]
\center{{\bf (a)}\hskip3.5cm{\bf (b)}}
\center{\psfig{figure=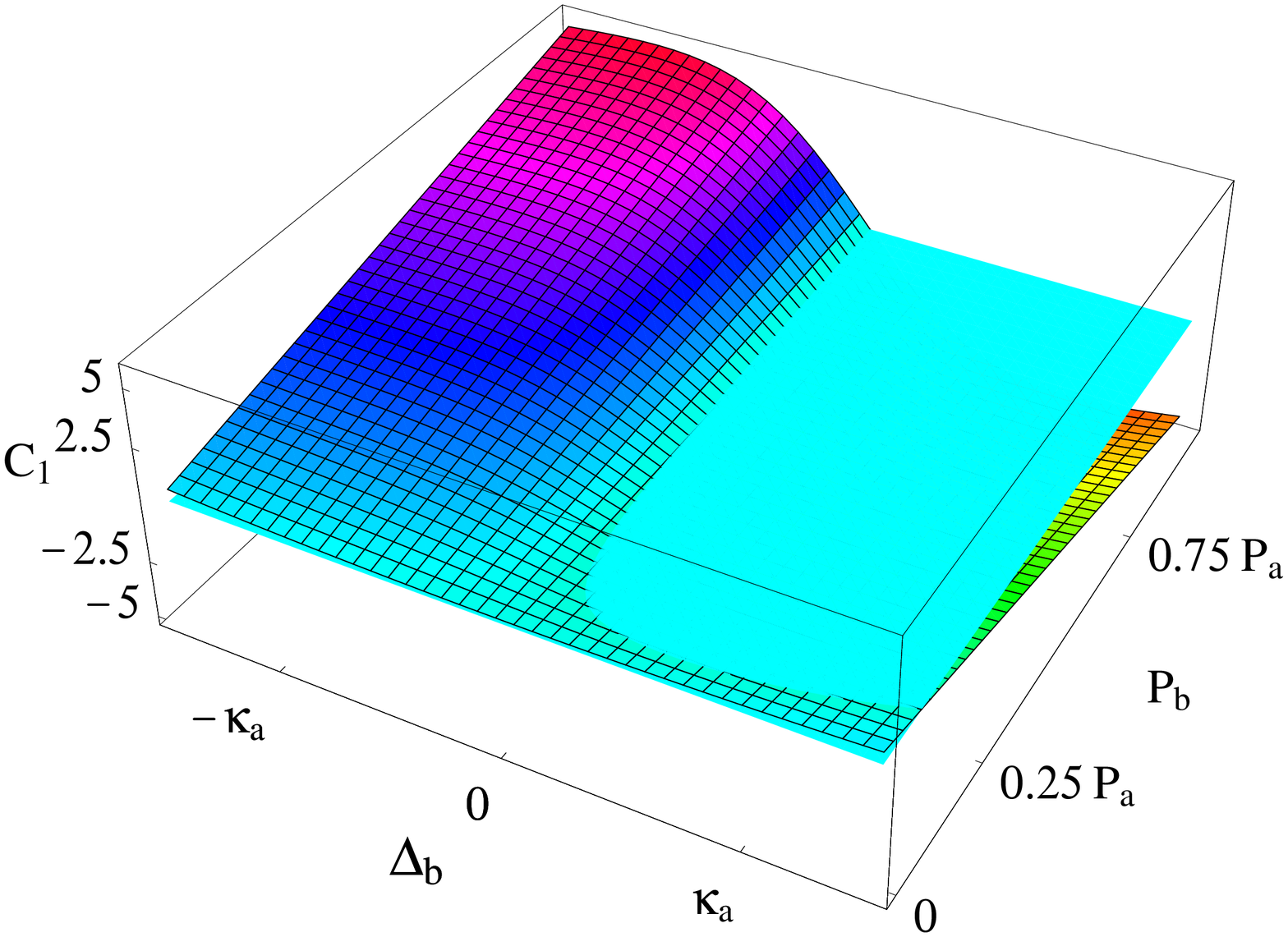,width=4cm,height=3.3cm}\,\,\psfig{figure=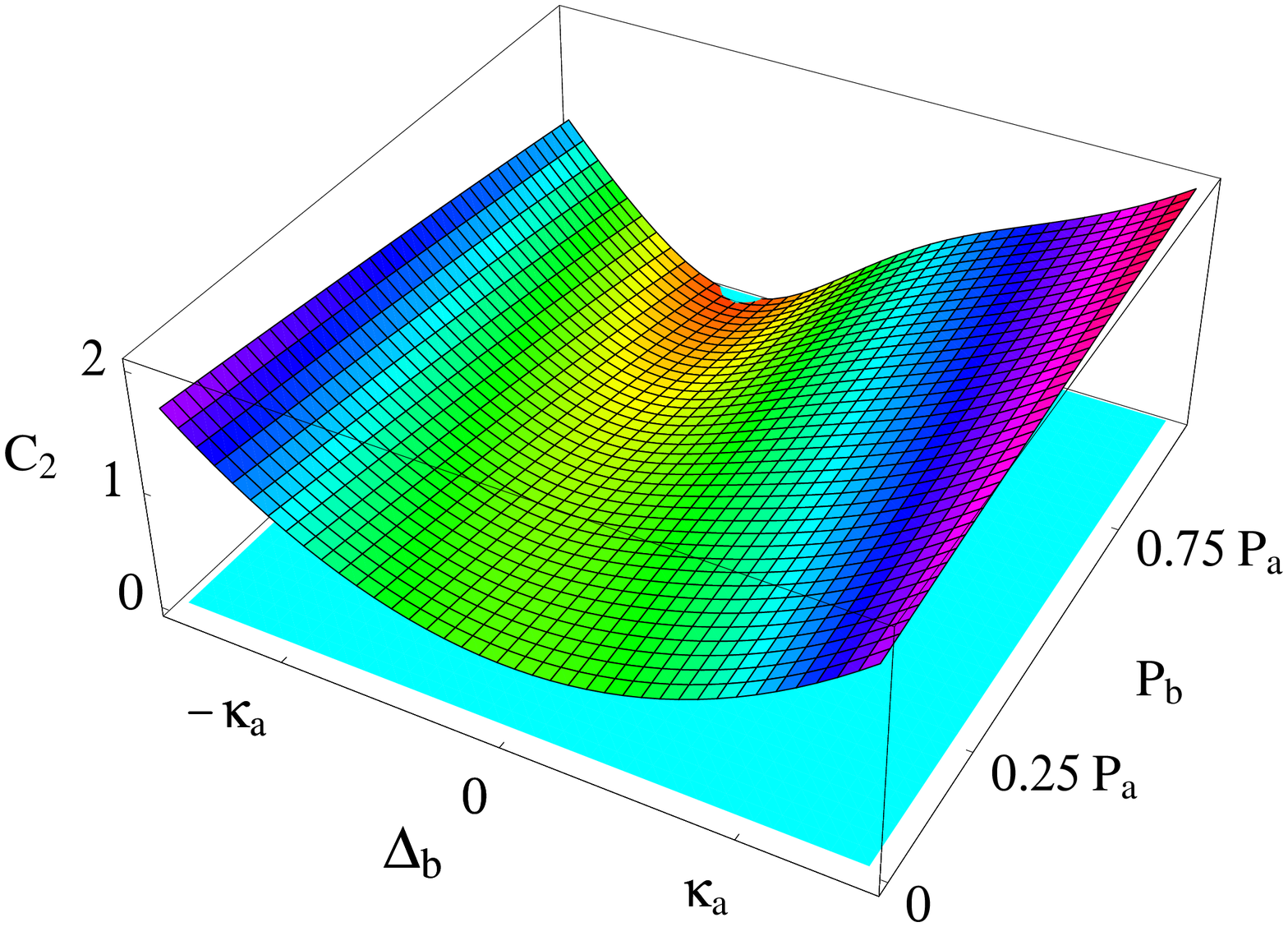,width=4cm,height=3.3cm}}
\caption{${C}_{1}$ (panel {\bf (a)}) and $C_2$ (panel {\bf (b)}) vs. $\Delta_b\in[-1.5\kappa_a,1.5\kappa_a]$ and $P_b\in[0,P_a]$ for $\Delta_a=\omega_m$. The horizontal plane corresponds to zero and is a help to the eye. We used $(\omega_m,\omega_{lj},\gamma_m,\kappa_j)/2\pi=(10^7,3.7\times10^{14},100,8.8\times10^7)$Hz, ${\ell}_j=1$mm, $T=0.4$K, $P_a=50$mW and $\mu=5$ng. 
}

\label{condT56}
\end{figure}
We 
assume the numbers in the caption of Figs.~\ref{condT56} 
for cavity $a$, which are very close to those of recently performed experiments on micromechanical systems~\cite{nature} and, to simplify the calculations, $\kappa_b=\kappa_a$ and $G_{0b}=G_{0a}$ (which can be easily relaxed). This allows us to study $C_{1,2}$ as functions of $\Delta_{b}$ and $P_{b}$. We take $\Delta_a=\omega_m$ as this choice corresponds to the maximum entanglement between $a$ and the mirror~\cite{david} and we conservatively assume that this holds also in presence of $b$ (which is a good approximation if $P_b\ll{P}_a$). With these choices, the behavior of $C_{1,2}$ is shown in 
Figs.~\ref{condT56}.
Even though for $P_b<{P}_a$ any sign of $\Delta_b$ corresponds to a stable regime, we focus on the region associated to $\Delta_b<0$ as we want to study the interaction of fields $a$ and $b$ with the mirror for any value of the back-action induced by $b$. 

{\it Intracavity entanglement. --}  At the steady state, 
$\hat{\bf f}(\infty)\equiv\!\hat{\bf f}_{ss}\!=\!\lim_{t\rightarrow\infty}\int^{t}_{0}d\tau{e}^{K\tau}\hat{\bf n}(t-\tau)$. The stationary covariance matrix $V_{pq}=\langle\hat{f}_{ss,p}\hat{f}_{ss,q}+\hat{f}_{ss,q}\hat{f}_{ss,p}\rangle/2$ of the tripartite system can be written as
\begin{equation}
\label{covariance3B}
{\mathbf V}=
\begin{pmatrix}
{\mathbf L}_{a}&{\mathbf C}_{ab}&{\mathbf C}_{am}\\
{\mathbf C}^{T}_{ab}&{\mathbf L}_{b}&{\mathbf C}_{bm}\\
{\mathbf C}^{T}_{am}&{\mathbf C}^{T}_{bm}&{\mathbf L}_{m}
\end{pmatrix}
\end{equation}
where ${\mathbf L}_j$ accounts for the local properties of subsystem $j=a,b,m$. ${\mathbf C}_{jk}$ 
describes the correlations between $j$ and $k$. The evaluation of ${\mathbf V}$ is performed using the Lyapunov equation 
${\mathbf V}\,{\mathbf K}+{\mathbf K}\,{\mathbf V}=-{\mathbf {N}}$,
which is found by noticing that, in the Markovian limit, ${\mathbf V}=\int^{\infty}_0{d}\tau({e}^{{\mathbf K}\tau}){\mathbf N}({e}^{{\mathbf K}\tau})^T$.
The Lyapunov equation is linear in the elements of ${\mathbf V}$, which can be easily determined, even though the formal solutions are cumbersome.  
\begin{figure}[t]
\center{\psfig{figure=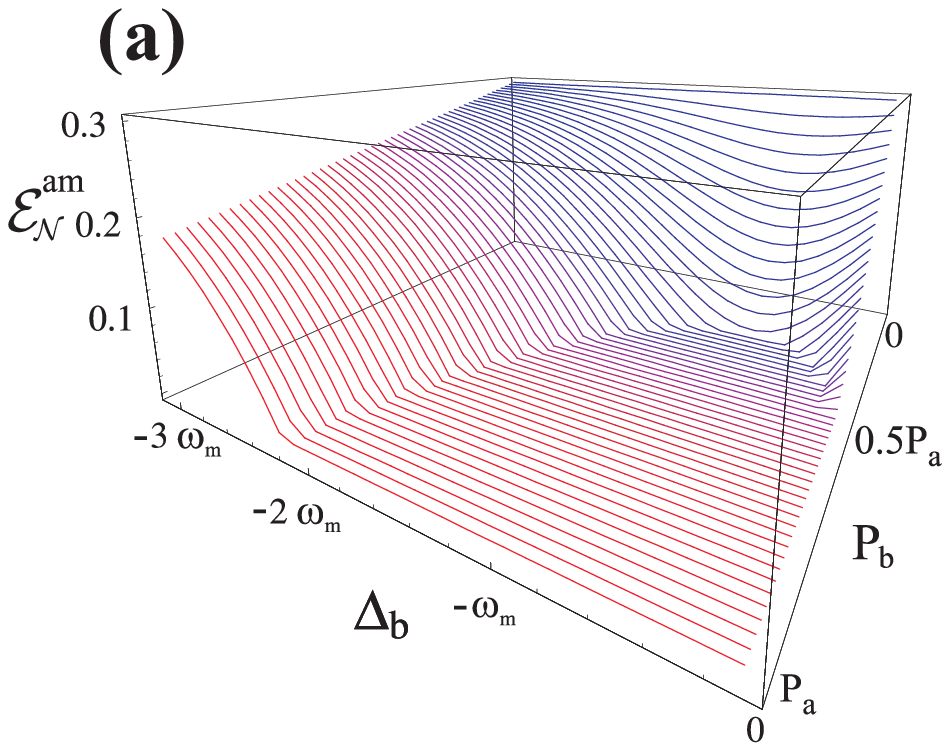,width=4.0cm,height=3.0cm}\psfig{figure=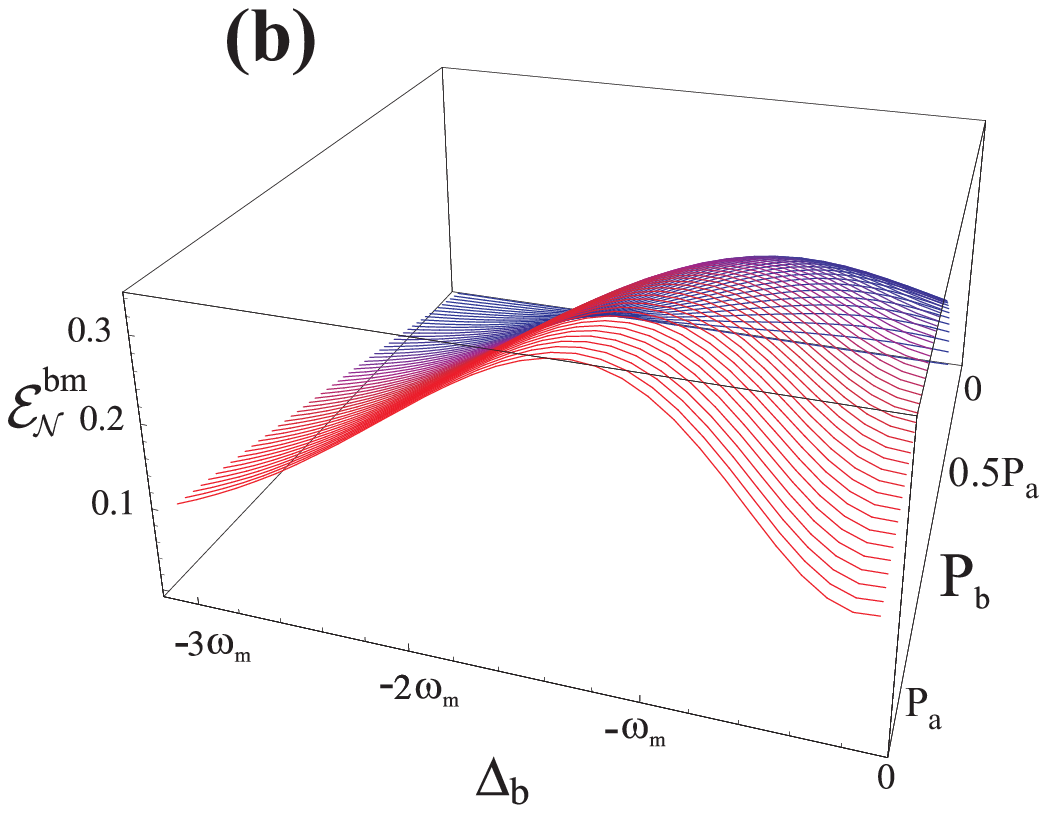,width=4.0cm,height=3.0cm}}
\caption{Logarithmic negativity ${\cal E}^{am}_{\cal N}$ (panel {\bf (a)}) and ${\cal E}^{bm}_{\cal N}$ (panel {\bf (b)}) vs.~$\Delta_b\in[-3\omega_m,0]$
for the parameters in the caption of Fig~\ref{condT56}. Each line corresponds to a specific value of $P_b\in[0,P_a]$. 
}
\label{ent3BmirrorPb}
\end{figure}

We can now study the behavior of the entanglement between the elements forming the tripartite system. In what follows, we characterize and quantify the bipartite entanglement in each intracavity field-mirror subsystem and in the field-field one. 
Quantitatively, we adopt the logarithmic negativity ${\cal E}^{jk}_{\cal N}$~\cite{logneg}, which is an entanglement monotone~\cite{logneg2} and can be calculated using the symplectic spectrum of the partially transposed reduced covariance matrix ${\mathbf V}^{{P}}_{jk}={\mathbf P}\,{\mathbf V}_{jk}\,{\mathbf P}$. Here, ${\mathbf V}_{jk}$ is the $4\times{4}$ submatrix extracted from ${\mathbf V}$ by considering the blocks in Eq.~(\ref{covariance3B}) relative to subsystems $j$ and $k$ only and ${\mathbf P}=\one_{2}\oplus{\bm \sigma}_z$ (with ${\bm\sigma}_r$ the $r$-Pauli matrix and $r=x,y,z$). The symplectic spectum $\{n_{\pm}\}$ is given by the eigenvalues of $|i{\bm \Sigma}{{\mathbf V}}^P_{jk}|$ with ${\bm \Sigma}=i{\bm \sigma}_{y}\oplus{i{\bm \sigma_y}}$
~\cite{alessio}. Explicitly
$(n^{\pm}_{jk})^2=[{\chi^-_{jk}\pm
({\chi^{-2}_{jk}-4\det{\mathbf V_{jk}}}})^{1/2}]/{2}$,
with $\chi^{\pm}_{jk}=\det{\mathbf L}_j+\det{\mathbf L}_k\pm2\det{\mathbf C}_{jk}$~\cite{alessio}. Entanglement in the state described by ${\mathbf V}_{jk}$ is found when $n^-_{jk}<1/2$, which translates the criterion for inseparability of Gaussian states (based on the negativity of the partial transposition criterion (NPT)~\cite{PPTsimon}) in the formalism of the symplectic spectrum. With these tools, ${\cal E}^{jk}_{\cal N}=\max[0,-\ln(2n^-_{jk})]$.

We start with the entanglement between field $a$ and mirror $m$. Without field $b$, the $a-m$ entanglement achieves its maximum~\cite{david}. However, the back action induced by the $b$ field could distort this picture, affecting the $a-m$ entanglement. A way to see this is to fix the working point for the $a$ cavity to those values corresponding to the maximum of entanglement with the mirror. Then, as the effects of the $b$ field are tuned, we study the behavior of the $a-m$ entanglement. To this task, as done before, we vary $\Delta_b$ and $P_b$ and 
examine the changes in entanglement from negligible to strong back action induced by $b$. The results are shown in Fig.~\ref{ent3BmirrorPb}.
\begin{figure}[ht]
\psfig{figure=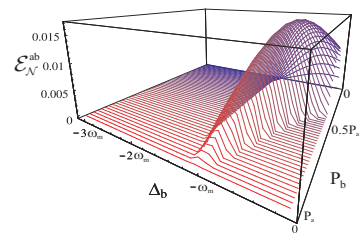,width=4.6cm,height=3.0cm}
\caption{Logarithmic negativity ${\cal E}^{ab}_{\cal N}$ vs. $\Delta_b\!\in[-3\omega_m,0]$ for 
the parameters in the caption of Fig.~\ref{condT56}. Each line corresponds to a value of $P_b\in[0,P_a]$. 
}
\label{ent3BcampiPb}
\end{figure}

\begin{figure}[t]
\psfig{figure=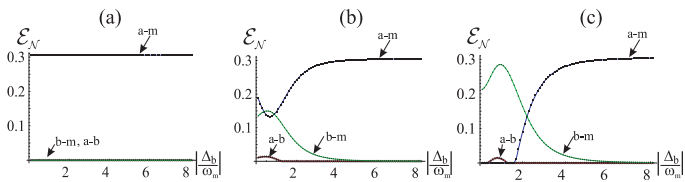,width=9cm,height=2.3cm}
\caption{${\cal E}^{jk}_{\cal N}$ vs. $|\Delta_b/\omega_m|\in[0,8]$ for increasing $P_b$. Panel {\bf (a)} is for $P_b=0$, {\bf (b)} for $P_b=0.15\,P_a$ and {\bf (c)} for $P_b\simeq{0.6}P_a$. Panel {\bf (b)} shows a situation where entanglement is found in any bipartite system obtained by tracing out one party.}
\label{sequence}
\end{figure}

As $P_b$ increases, the back-action of field $b$ on $m$ becomes more relevant, thus affecting the $a-m$ entanglement at moderate values of $\Delta_b$.
However, if $\Delta_b$ grows, ${\cal E}^{am}_{\cal N}$ revives achieving again values close to its maximum (which is larger than $0.3$~\cite{david}), even more evidently if the range of $\Delta_b$ is increased as in Fig.~\ref{sequence}. Indeed, in a far-detuned cavity, less input power enters thus taking back the system to a situation of small back-action. The reduction in ${\cal E}^{am}_{\cal N}$ is caused by a simultaneous raise of ${\cal E}^{bm}_{\cal N}$, as shown in Fig.~\ref{ent3BmirrorPb} {\bf (b)}. Indeed, the calculation of ${\cal E}^{bm}_{\cal N}$ reveals that entanglement is established between the two subsystems, pronounced in the region of moderate $\Delta_b$ where ${\cal E}^{am}_{\cal N}$ suffered the effects of $b$'s back-action. In this case, large detunings lower ${\cal E}^{bm}_{\cal N}$ which, eventually, goes to zero as $\Delta_b\gg{\omega_m}$ (Fig.~\ref{sequence}).

The complementary behavior of ${\cal E}^{jm}_{\cal N}$'s is an evidence of the way the presence of $b$ enables us to infer the features of system $a-m$: the ancillary field $b$ gets 
entangled with $a$ and $m$, at the expense of the ${\cal E}^{am}_{\cal N}$. Given the symmetry between $a$ and $b$, this same claim holds if we swap $a$ and $b$. Indeed, the most interesting aspect of the entanglement dynamics comes from the study of ${\cal E}^{ab}_{\cal N}$ (see
Fig.~\ref{ent3BcampiPb}). As soon as ${\cal E}^{bm}_{\cal N}$ is established, ${\cal E}^{ab}_{\cal N}$ becomes non-zero. 
By comparing the results obtained for increasing 
$P_b$ 
and $\Delta_b$ fixed at the value corresponding to the maximum of ${\cal E}^{ab}_{\cal N}$, we see that ${\cal E}^{ab}_{\cal N}$ disappears, slowly with respect to ${\cal E}^{am}_{\cal N}$ (at $P_b\simeq{P_a}$, ${\cal E}^{am}_{\cal N}=0$ and
${\cal E}^{ab}_{\cal N}\neq{0}$). 

We may use the entanglement between $a$ and $b$ as a 
tool to see signatures of entanglement between $a$ and $m$, 
as $a$ and $b$ never directly interact and all ${\cal E}^{ab}_{\cal N}\neq{0}$ is necessarily due to a mediation by the mirror. That is, in this system the mirror acts as a bus for the cross-talking of the fields. Any entanglement between $a$ and $b$ is thus an indication of a coherent field-field interaction through radiation pressure. 
As the input fields are prepared in pure
coherent states, ${\cal E}^{ab}_{\cal N}$ must be the result of an effective entangling field-field interaction~\cite{myung}. Moreover, 
by finding ${\cal E}^{ab}_{\cal N}\neq{0}$ we can infer entanglement between one of the fields and the mirror, even though the converse is not true (there are situations where ${\cal E}^{ab}_{\cal N}=0$ with ${\cal E}^{jm}_{\cal N}\neq{0}$, as in Fig.~\ref{sequence} {\bf (a)}).  

Therefore, ${\cal E}^{ab}_{\cal N}$ can be taken as a signature of entanglement between the cavity $a$ and the mirror $m$. If initially, $a$ and $b$ are in pure states, ${\cal E}^{ab}_{\cal N}>0$ strictly indicates entanglement between $a-m$. While, in general, one could construct models where two systems become entangled via the coupling to a system that remains separable with respect to the rest~\cite{cubitt}, the interaction studied in our indirect scheme provides strong evidence for mirror-cavity entanglement. The case of Fig.~\ref{sequence} {\bf (b)} is interesting: for $\Delta_b/\omega_m\simeq{0.5}$, ${\cal E}^{am}_{\cal N}\simeq{\cal E}^{bm}_{\cal N}$ with ${\cal E}^{ab}_{\cal N}$ achieving its maximum, thus {\it optimizing} the overall entanglement distribution within the system. By means of NPT we have checked that, in these conditions, genuine tripartite entanglement is shared between the subsystems. The study of an entanglement monogamy inequality in our system and lower
bounds exploiting a promise to the interaction 
will be the focus of further investigations~\cite{commentodopo}.


{\it Extracavity description. --} Even though the entanglement between the intracavity fields and the mirror is the object of our investigation, the accessible quantities in this system are given by the fields leaking out of the cavities. We now show a simple operative strategy to infer the correlation properties of the intracavity system. We aim at estimating the field-field covariance matrix at the output. We assume $\kappa_j=\kappa$ (the generalization is straightforward) and define $\hat{f}^{in}_{p}(t)=\hat{n}_{p}/\sqrt{2\kappa}\,\,(p=1,..,4)$. The extracavity field quadratures $\hat{\mathbf f}^{out}=(\delta\hat{x}^{out}_a,\delta\hat{y}^{out}_a,\delta\hat{x}^{out}_b,\delta\hat{y}^{out}_b)$ are related to the intracavity ones by the input-output relations $\hat{f}^{out}_{p}(t)=\sqrt{2\kappa}\hat{f}_p(t)-\hat{f}^{in}_p(t)$~\cite{collett}. The outputs are free fields and their dimension is {sec}$^{-1/2}$. It is thus convenient to introduce dimensionless extracavity quadratures which we use to build up the output covariance matrix. One way is to define $\hat{f}^{\nu}_{d,p}=\lim_{t\rightarrow\infty}\frac{1}{\sqrt{t_m}}\int^{t+t_m}_{t}\hat{f}^{\nu}_p(t')dt',\,\,(\nu=in,out)$, where $t_m$ is the {measurement time}, i.e., the acquisition time chosen for a measurement of the output quadratures at the steady state. It is easy to check that $\hat{f}^{\nu}_{d,p}$'s satisfy the usual canonical commutation rules. In this way, the input-output relations become $\hat{f}^{out}_{d,p}=\sqrt{{2\kappa{t_m}}}\hat{f}_{ss,p}-\hat{f}^{in}_{d,p}$, where we used $\hat{f}_p(\infty)=\hat{f}_{ss,p}$. 

With this notation, $V^{out}_{pq}=\langle\hat{f}^{out}_{d,p}\hat{f}^{out}_{d,q}+\hat{f}^{out}_{d,q}\hat{f}^{out}_{d,p}\rangle/2$ is easily evaluated. We find ${\mathbf V}^{out}=2\kappa{t}_{m}{\mathbf V}_{ab}+{\mathbf V}^{in}$, where $V^{in}_{pq}=\frac{1}{2}\delta_{pq}$. 
The simplicity of the expression relating the extracavity correlations to the analogous intracavity quantities suggests an operative way to infer the entanglement behavior of fields $a$ and $b$. For a fixed working point and a value for $t_m$ (typically $\sim{1/\kappa}$), ${\mathbf V}^{out}$ is built up by homodyne measurements~\cite{david,myungmunro}. Then, ${\mathbf V}_{ab}$ can be reconstructed as $({\mathbf V}^{out}-{\mathbf V}^{in})/2\kappa{t}_m$. This prescription is just an additional step in the numerical postprocessing of the data required for the estimation of the entanglement. 

{\it Conclusions. --} We have introduced a scheme to reveal entanglement between a cavity field and a movable mirror by inducing quantum correlations in the tripartite system which includes an ancillary field. Using state of the art parameters, we have studied the effects of back-action by the ancilla on the entanglement which has to be inferred. We found a working point at which entanglement appears in any bipartite subsystem. 
Present work includes the 
formulation of a lower bound to optomechanical entanglement related to the detected optical one, which will allow for the use of field-field correlation as a quantitative witness for the mirror-field entanglement. We hope that the 
presented work will pave the way to the experimental inference of entanglement involving macroscopic objects.

{\it Acknowledgements. --}  We thank F. Blaser, J.\ Kofler and A. Zeilinger for discussions. We acknowledge support by the Austrian Science Fund, KRF, the UK EPSRC, the EU  under the Integrated Project Qubit Applications QAP funded by the IST Directorate (contract number 015846), Microsoft Research, and the EURYI Award Scheme. MP acknowledges The Leverhulme Trust for financial support (ECF/40157).


\begin{thebibliography} {99}
\bibitem{commentophase} We can choose the phase reference of the cavity fields so that $\alpha_{s,j}$'s are real~\cite{david}.
\end{thebibliography}
\end{document}